\documentclass[preprint,showpacs,preprintnumbers,amsmath,amssymb]{revtex4-1}
\usepackage{graphicx}
\usepackage{bm}
\usepackage{color}
\usepackage{amssymb}
\usepackage{epsfig}
\usepackage{amsmath}
\begin{document}

\title{Statistical analysis of chiral structured ensembles: role of matrix constraints}
\author{Triparna Mondal and  Pragya Shukla  }
\affiliation{Department of Physics, Indian Institute of Technology,Kharagpur, India}
\date{\today}


\begin{abstract} 	

	We  numerically analyze the statistical properties of complex system with conditions subjecting the matrix elements to a set of specific constraints besides symmetry, resulting in  various structures in their matrix representation.   Our results reveal an  important trend: while  the spectral statistics is strongly sensitive to the number of independent matrix elements, the eigenfunction statistics seems to be affected only by their relative strengths. This is contrary to previously held belief of one to one relation between the statistics of the eigenfunctions and eigenvalues (e.g. associating Poisson statistics  to the localized eigenfunctions and Wigner-Dyson statistics to delocalized ones).

\end{abstract}

                                                 
\maketitle

\section{Introduction}

The missing information due to complexity in a system manifests itself by full/ partial randomization of the matrix representations of the operators. The statistical behavior of the complex system can then be described by an appropriate random matrix ensemble  taking into account the system conditions.  The conditions that influence  nature of the ensemble  can be divided into two types: (i) the ``matrix" constraints (e.g., conservation laws and symmetry) which affect the broad structure of a single matrix through transformation properties and collective relations among the elements, and, (ii) the ``ensemble" constraints (e.g., disorder, dimensionality and boundary conditions) which manifest themselves through the  ensemble parameters, i.e., the distribution properties of matrix elements and/or the local relations among them \cite{psijmp}.  In past there have many studies of the ensembles with the matrix consitarints related to unitary and anti-unitary symmetries \cite{fh, me, brody}.
But information about the ensembles with matrix constraints based on conservation laws, which lead to specific relations among matrix elements, is still missing. The appearance of such cases in wide range of complex systems e.g disordered systems \cite{chal, mm, pstrans}, complex and neural networks \cite{ggs,qw,afm,ahn}, financial markets \cite{cmz}  make their statistical studies highly desirable. This motivates the present study in which we seek and analyze those ``matrix'' constraints which may affect the eigenfunction localization (spread of eigenfunctions in the basis-space) in a way similar to the influence of disorder (which is an ensemble constraint). Our primary focus is to understand the connection between degree of localization of eigenfunctions and the nature of spectral statistics.

The symmetry is a matrix constraint playing a very important role in physical properties \cite{dy, zirn, fh}. Based on nature of the symmetry operators e.g. continuous or discrete,  unitary, anti-unitary or their combinations, the generators of the dynamics can be divided into various universality classes \cite{zirn}. The role of sysmmetry  on statistical properties of the spectrum and eigenfunction dynamics has been studied extensively in past. The presence of conservation laws along with symmetries however may lead to new structures in the matrix. Although a special class of structures have been studied in recent past \cite{grela, by, ossi},  the information about the role of generic structures in the eigenfunction localization is still missing.  In the present work, we attempt to fulfil this gap by a numerical analysis of some  structured matrix ensembles.

Based on the nature of  constraints on its elements, a matrix can display a range (host) of structures. Such structured matrices e.g. circulant, toeplitz, column/row constrained not only appear in many areas of physics \cite{qw,ach,cd,grela,ossi}, they have recently been proposed to be useful for reduction of the computational complexity of large-scale neural networks \cite{qw, afm, ahn,zlwlty}. One of the most successful machine learning strategies, nowadays, is deep neural network algorithms which is being used for speech recognition, computer vision, and classification of large data sets. In deep non-negative matrix factorization (NMF), a topic model for data representation and feature extraction, the untructured weight matrix can be compressed to structured matrix without losing much accuracy while achieving high compression ratio and high speedup for a broad category of network models \cite{zlwlty}. As various properties of such networks can at least in principle be described in terms of the eigenfunctions and eigenvalues of the related matrix, seeking more information about the latter is highly desirable. This motivates us to analyze the ensembles of a range of such structured matrices. As our primary focus in this work is to seek the effect of matrix constraints on the eigenfunction localization, we confine our study to the cases with different degree of localization.

The dimensionality of the system is an important matrix constraint which governs the sparsity of the matrix \cite{psand, psijmp}. This alongwith the ensemble constraints (e.g. type of randomness of the matrix elements) plays a significant role in eigenfunction localization. For a specific set of matrix constraints, it is well known that a localization to delocaliztaion transition, (referred as LD transition hereafter) of the eigenfunctions can be brought about by varying one of the local constraints e.g disorder; a well-known example in this context is Anderson transition \cite{psand}.  Here we consider the reverse question: can the above transition be brought about by varying matrix constraints for a fixed set of local constraints (while keeping the symmetries invariant to keep the   transformation class of the ensemble unchanged)?
For our analysis we consider chiral ensemble of structured matrices. This not only helps us to analyze the role of this symmetry along with other matrix constraints but also preserves the Hermitian nature of the matrices which is relevant for application to a wider range of real physical  
systems \cite{chal}.

The paper is organized as follows. We consider five different matrix constraints on chiral matrix while keeping the  ensemble parameters same. Section II briefly describes these cases along with their ensemble densities.  The later in principle can be used to derive the joint probability distribution of the eigenvalues and eigenfunctions and thereby various fluctuation measures.  The constraints however lead to correlations among matrix elements which makes an exact derivation of related statistical measures technically difficult and approximations are necessary. An insight into the statistical behavior however can be gained by the numerical analysis; this is  presented in section III.  We conclude in section IV with a review of our main results and open questions.


\section{Hermitian Matrices with chirality and other constraints}

A generic $2N\times 2N$ chiral Hamiltonian H is given by
\begin{eqnarray}
H= \left( {\begin{array}{cc}   0  & C  \\  C^{\dagger} & 0 \end{array} } \right).
\label{ch1}
\end{eqnarray}
where $C$ is a general $N\times(N+\nu)$ real or complex or quaternion matrix, (depending on the nature of exact anti-unitary symmetry of $H$). Both the spectral and eigenfunction statistics of $H$ matrix depend on the nature of $C$. For $C$ subjected to Hermitian constraint only, its bulk spectral as well as eigenfunction correlations can be modeled by the Wigner-Dyson universality classes. In presence of chirality, however an additional level repulsion appears around zero which leads to different spectral correlations near zero (the origin) and away from the bulk \cite{evan}.

For a simple exposition of the influence of constraints on the eigenvalues and eigenfunctions, we choose $C$ as a real non-symmetric square matrix with $\Lambda$ as its eigenvalue matrix  and $U, V$ as the left and right eigenvector matrices  respectively:  $C =U \Lambda V$ with $U . V=I$ and $I$ as the identity matrix. 
The left and right eigenvectors $U_n, V_n$ ($n^{th}$ row of $U$ and $n^{th}$ column of $V$ respectively) of $C$, corresponding to the eigenvalue $\lambda_n$, are then given by following relation: 
\begin{eqnarray}
U_n \; C =  \lambda_n \; U_n, \hspace{0.3in} C \; V_n = V_n \; \lambda_n.
\label{cuv}
\end{eqnarray}

With $H$ given by eq.(\ref{ch1}), let $E$ be the eigenvalue matrix ($E_{mn} =e_n \delta_{mn}$) and $O$ as the eigenvector matrix of $H$, with $O_{kn}$ as the $k^{th}$ component of the eigenvector $O_n$ corresponding to eigenvalue $e_n$. The above along with  eq.(\ref{ch1}) implies that the  eigenvalues  of $H$ exist in equal and opposite pairs; let us refer such pairs as ${e}_n, e_{n+N}$ with $e_n=- e_{n+N}$, $1\le n \le N$.  
The eigenvector pair $O_n, O_{n+N}$ corresponding to eigenvalue pair $e_n, e_{n+N}$ can in general be written as $\left(\begin{array}{cc} X_n  \\ \pm Y_n \end{array}\right)$ with $X_n, Y_n$ as column vectors with $N$ real components.  Eq.(\ref{ch1}) then gives $C \; Y_n = e_n \; X_n$ and $C^{\dagger} \;  X_n = e_n \; Y_n$ which leads to $C^{\dagger} C \; Y_n = e_n^2 \; Y_n$ and $C C^{\dagger} \; X_n = e_n^2 \; X_n$.
Further note the orthogonality condition $O_n^{\dagger}.O_{n+N} =0$ along with normalization $O_n^{\dagger}.O_n=1$ gives $X_n^{\dagger}. X_n=Y_n^{\dagger}. Y_n =1/2$.

With $C$ as a $N\times N$ square matrix with real elements, it has $N^2$ free parameters.  Introduction of new  ''matrix" constraints results in correlations among the matrix elements and reduces the number of free parameters, latter referred as $M$ hereafter. Here we consider following  five cases of the matrix $C$, given in a sequence of decreasing number of constraints (i.e with increasing number of free parameters):

\textbf{Case 1: \textit{Column-constraint Circulant matrix:}}  For  $C$ as a $N\times N$ real, circulant matrix \cite{toep} with its elements 
$C_{kl} = c_{(k-l) \; mod \; N}$ , for $k,l=1\rightarrow N$
the number of independent elements $M=N$. Further imposing the column(row)  constraints  i.e.  
\begin{eqnarray}
\sum_{k=1}^{N} C_{kl} = \sum_{k=1}^{N} C_{lk} = \alpha 
\label{cc}
\end{eqnarray}
with $\alpha$ as a real constant and same for each column and row, $M$ is further reduced: $M=N-1$. For $C$ as a circulant matrix,  the right and left eigenvector matrices are same $U=V$ and both  $C, C^{\dagger}$ have same set of eigenvectors. This in turn implies $U_n$ as the eigenvector of $C C^{\dagger} = C^{\dagger} C$ with eigenvalue $|\lambda_n|^2$. Further as $\lambda_n = \lambda_{N-n}^*$ for $n < N$, and the eigenvalue pairs of H matrix $e_n,e_{n+N}$ appears with $e_n=|\lambda_n|, e_{n+N}=-|\lambda_n|$. This gives
$X_n = Y_n = \eta \; \left(U_n + (1-\delta_{nN}) \; U_{N-n} \right)$
where the real constant $\eta$ can be determined by orthogonality condition on $O_n$.

As clear from the above alongwith  eq.(\ref{ch1}), $H$ in this case has four matrix constraints (i) chiral symmtery,  (ii) hermiticity, (iii) circulant constraint, (iv) column row constraint.  Herafter this case will be referred as the column-constraint chiral matrix with circulant off diagonal blocks. The spectral properties of this case was considered in detail in \cite{tss} and is included here for comparison with other cases. 
As discussed in \cite{tss},  all eigenvectors of $H$ (i.e $O_n$, $n=1, \ldots, 2N)$ in this case remain extended, with their IPR given by $I_2(O_n) = {3 \over 4N}$ for $n \not=N, 2N$ and $I_2(O_N)=I_2(O_{2N}) ={1\over  2N}$. For the case $\alpha=0$, however $\lambda_N=0$ leading to a degenerate pair $e_N, e_{2N}=0$ with corresponding   eigenvectors as $\left(\begin{array}{cc} U_N  \\ 0 \end{array}\right)$ and $\left(\begin{array}{cc} 0  \\  U_N \end{array}\right)$. As a consequence 
$I_2(O_N) =I_2(O_{2N}) = 1/N$ for the case $\alpha=0$.

\textbf{Case 2: \textit{Toeplitz matrix:}} 
Next we consider  $C$ as a $N \times N$ toeplitz matrix with real elements \cite{toep}, defined as
\begin{eqnarray}
C_{kl} = C_{(k+1),(l+1)} = c_{(k-l)}
\label{toep}
\end{eqnarray}
and $H$ now becomes a $2N\times 2N$  chiral matrix with Toeplitz off diagonal blocks.

The absence of circulant constraint as well as column constraint in $C$ reduces the correlations among its matrix elements and increases the number of its independent parameters $M=2N-1$.  Based on previous studies, some information is available for the eigenvectors and eigenvalues of these matrices \cite{toep}. Contrary to case 1, however a general formulation of the eigenvalues and eigenvectors of these matrices is technically difficult. Eq.(\ref{toep}) however in general implies

\begin{eqnarray}
\sum_{n=1}^N \lambda_n \; \left( U_{nk} \; V_{nk+r} - U_{nl} \; V_{n l+r}\right) =0 \hspace{0.1in} \forall (k-l)=r
\label{toep1}
\end{eqnarray}
The above relation being valid for any $(k,l)$ pair at a fixed distance $r=k-l$, this clearly indicates 
strong correlations among all the eigenfunctions of $C$ and only one of them, say $U_1$ is independent. 
This again affects the Jacobian of transformation, reducing the level-repulsion significantly.

\textbf{Case 3: \textit{Column constraint matrix with same diagonals: }}  $C$ is now obtained by relaxing the Toeplitz constraint and by imposing the column (row) constraint along with the condition that all diagonal elements in the matrix $C$ are equal.  The diagonals $C_{kk}$ can be written as 
\begin{eqnarray}
C_{kk} = C_{11} = \alpha-\sum_{n=2}^{N}C_{1n} =\alpha-\sum_{n=2}^{N}C_{n1}  
\quad \quad {\rm for} \quad k=1,\dots,N.
\label{dg}
\end{eqnarray}
The off-diagonals of $C$ are randomly chosen while keeping the sum of those in a column or row as constant. This increases the number of independent elements in $C$ with $M=N^2-3N+1$. 
Taking $\alpha=0$ and using the relation $\sum_{k=1}^N H_{kk} = \sum_{n=1}^N e_n$, eq.(\ref{dg}) can be rewritten as the condition on a combination of the eigenvalues and eigenfunctions:
\begin{eqnarray}
\sum_{n=1}^{N} \; \lambda_n \left(1-N U_{nk} V_{nk} \right) = 0
\quad \quad {\rm for} \quad k=1,\dots,N.
\label{dg1}
\end{eqnarray}
Clearly the components of the right and left eigenvector of a specific eigenvalue say $\lambda_k$ are not independent. The decreased number of constraints are expected to 
help delocalization of the eigenfunctions; this is confirmed by our numerics given in section IV.

\textbf{Case 4: \textit{ Upper Toeplitz matrix:}} Another form of $C$ can be obtained by  removing column as well as diagonal constraints from $C$ but imposing a variant of the Toeplitz  constraint. The latter corresponds to the Toeplitz condition only among the elements of the diagonal and the upper diagonal of $C$:
\begin{eqnarray}
C_{k,l} = C_{(k+1),(l+1)} = c_{(k-l)} \quad \quad {\rm for} \; \; {k\leqslant l} 
\label{th1}
\end{eqnarray}
As no constraint is imposed on the lower diagonals, these are independent from each other as well from those in upper diagonal. With chiral symmetry in $H$, the number of independent elements in this case become $M=\frac{1}{2}[N(N+1)]$ (Note here $C$ is $not$ a symmetric matrix). Eq.(\ref{toep1}) is still valid in this case but only for $k \leqslant l$.

\textbf{Case 5: \textit{Column-constraint matrix}} A column-constraint chiral matrix with no other constraints (global) associated with it has block matrix $C_{N\times N}$ satisfying one constraint (\ref{cc}). 
The number of free parameters in $C$ are now $N^2-2N+1$. The case with $C$ as a real-symmetric matrix was considered in detail in \cite{ss1,ss2}. As number of constraints in the present case are minimum, the eigenfunction are expected be less localized as compared to the cases (1-4).

\section{Ensembles with chriality and other constraints}

For matrices representing complex systems, it is imperative to consider their ensembles which can subsequently be used to derive the joint probability distribution (JPDF)  of its eigenvalues and eigenfunctions and other related properties. Consider the ensemble density $\rho(H)$ of  $H$. Following from eq.(\ref{ch1}), 
\begin{eqnarray}
\rho(H) = J_c(H | C) \; \rho_c(C)
\label{rhoh}
\end{eqnarray}
 with $J_c(H|C)$ as the Jacobian of transformation from $C$-space to $H$-space and $\rho_c(C)$ as the probability density of the ensemble of $C$ matrices.
For cases where $\rho_c(C)$ is not known, one can invoke the  maximum entropy hypothesis: the system is best described by the distribution $\rho_c(C)$ that maximizes Shannon's information entropy
$I[\rho_c(C)] = - \int \rho_c(C)\; {\rm ln} \rho_c(C)\; {\rm d}\mu(C)$
under known set of ensemble constraints e.g. on the moments of entries of $C$. It must be emphasized that a specific set of matrix constraints can lead to many different ensembles. 

For simple exposition of our ideas, here we consider the set of ensemble constraints  which leads to a Gaussian distribution for all independent parameters $c_{\mu}$ of $C$, with $\mu=1, \ldots, M$ each with same variance and zero mean: 
\begin{eqnarray}
\rho_c(C) =  \mathcal{N} \;  {\rm exp}\left[- {1 \over 2 \sigma^2} \; \sum_{\mu=1}^M    \; c_{\mu}^2   \right]  \; F_c 
\label{rhoc}
\end{eqnarray}
with $\mathcal{N} $ as a normalization constant, $M$ as the total number of independent parameters in $C$  and the function $F_c$ describes the set of matrix constraints on $C$. For the cases 1-5 mentioned in previous section, $\rho_c(c)$ can be written more explicitly as follows.

\textbf{Case 1:} With only $(N-1)$ free parameters, the  probability density of $C$ in this case becomes :
\begin{eqnarray}
\rho_c(C) &=&  \mathcal{N} \;  {\rm exp}\left[-   {1\over 2 N \sigma^2}   \sum_{k,l=1}^N \; \mit | C_{kl} \mit |^2 \right]  \; F_1
\label{rhoc1}
\end{eqnarray}
where the function $F_c$ gives the circulant as well as column/row constraint:
\begin{eqnarray}
 F_1  \equiv \delta \left(\sum_{l=1}^N  C_{1l} - \alpha \right) \; \prod_{k,l=1}^N \delta(C_{kl} - c_{(k-l) \; mod \; N}). 
\label{fc}
\end{eqnarray}

\textbf{Case 2:} 
With number of independent parameters now increased to $2N-1$,  the ensemble density of $C$ can be expressed as
\begin{eqnarray}
\rho(C) =  \mathcal{N} \;  {\rm exp}\left[-   {1\over 2 \sigma^2} \sum_{k, l=1}^N   \alpha_{kl}  \; | C_{kl} |^2   \right]  \; F_2(C)
\label{rhot1}
\end{eqnarray}
with 
\begin{eqnarray}
\alpha_{kl}= {1\over N-|k-l|}
\label{akl2}
\end{eqnarray} 
and $F_c$ now  refers to  Toeplitz constraint only:
\begin{eqnarray}
 F_2(C) \equiv \prod_{k,l=1}^N \delta(C_{kl} - c_{(k-l)}). 
\label{fc2}
\end{eqnarray}

\textbf{Case 3:} Under the constraints of equal diagonal elements  along with fixed column (and row) sums,  the distribution of $C$ matrix can be written as
\begin{eqnarray}
\rho(C) &=& \mathcal{N} \; {\rm exp}\left[- {1\over 2 \sigma^2} \left(C_{11}^2 + \sum_{k,l=1; k \not=l}^{N-1}  C_{kl}^2 \right) \right]  \; F_3 \\
&=&  \mathcal{N} \; {\rm exp}\left[- {1\over 2 N \sigma^2 } \sum_{k=1}^{N-1} C_{kk}^2 -{1\over 2 \sigma^2} \sum_{k,l=1; k \not=l}^{N-1}  C_{kl}^2 \right]  \; F_3 
\label{rhod1}
\end{eqnarray}
with
\begin{eqnarray}
F_3 \equiv \delta \left(\sum_{k=1}^N C_{k1}-\alpha\right) \prod_{l=2}^N \delta \left(\sum_{k=1,k\not = l}^N  C_{kl} - (C_{kk}+\alpha) \right) .
\label{fc3}
\end{eqnarray}

Proceeding as in previous case, $\rho_{c}$ can again be rewritten as eq.(\ref{rhot1}) but with following changes:
\begin{eqnarray}
\alpha_{kk} ={1\over N}, \qquad 
\alpha_{kl}=1, \qquad \alpha_{kN}=\alpha_{Nl}=0 \quad {\rm for} \; k, l <N
\label{akl3}
\end{eqnarray}

\textbf{Case 4:} Again reducing all the other constraints from block matrix $C$, a correlation is introduced among the elements of the diagonal and the upper triangle of $C_{N\times N}$:
\begin{eqnarray}
C_{k,l} = C_{(k+1),(l+1)} = c_{(k-l)} \quad \quad {\rm either} \; k\leqslant l 
\label{diag1}
\end{eqnarray}
along with other elements being random in the lower off-diagonals. 
\begin{eqnarray}
\rho_c(C) =  \mathcal{N} \;  {\rm exp}\left[-   {1\over 2 \sigma^2} \left( \sum_{k,l=1; k> l}^N   \; C_{kl}^2   +  \sum_{k,l=1; k \leqslant l}^N   {1\over N-|k-l|}  \; C_{kl}^2  \right)\right]  \; F_4
\label{rhoc2}
\end{eqnarray}
with
\begin{eqnarray}
 F_4\equiv \prod_{k,l=1;k\geq l}^N \delta(C_{kl} - c_{(k-l)}). 
\label{fc4}
\end{eqnarray}

Following case 2, $\rho_c$ can be rewritten as eq.(\ref{rhot1}) but with $\alpha_{kl}$ now given as follows: 
\begin{eqnarray}
\alpha_{kl} =1 \quad {\rm for} \; k > l, \qquad 
\alpha_{kl} = {1\over N-|k-l|} \quad {\rm for} \; k \leqslant l.
\label{akl4}
\end{eqnarray}

\textbf{Case 5:} A column-constraint chiral ensemble with no other constraints (global) associated with it has block matrix $C_{N\times N}$ satisfying one constraint (\ref{cc}) only with most of the elements random, i.e., with distribution (\ref{rhoc}) along with 
\begin{eqnarray}
F_5 \equiv \prod_{l=1}^{N} \delta \left(\sum_{k=1}^N C_{kl}-\alpha\right). 
\label{fc5}
\end{eqnarray}

It is worth mentioning here that the  Gaussian form of $\rho_c$ in eq.(\ref{rhoc}) is a consequence of the conditions on  the $1^{\rm st}$ and  $2^{nd}$ order moments of $c_{\mu}$.  Higher order moments of the latter can be subjected to similar constraints too which would lead to non-Gaussian ensembles of chiral constrained matrices. A most generic form of $\rho_c$ can be given in terms of the JPDF of circulant variables $c_j$, with $j=0 \to M$: $\rho_c(C) =  \mathcal{N} \; \rho_0(c_1, c_2, \ldots, c_M)  \; F_c$, with $F_c$ dependent on matrix constraints.  The present work however is confined to the study to the ensembles given by eqs.(\ref{rhoc1}, \ref{rhot1}, \ref{rhod1}, \ref{rhoc2}) and eq.(\ref{rhoc}) along with eq.(\ref{fc5}); these will be used in next section for the numerical statistical analysis of the cases $1-5$, respectively.

\section{Numerical analysis}

In principle,  the eigenvalue and eigenfunction distributions can be derived from the  ensemble density $\rho(H)$. This however requires a knowledge of the Jacobian of transformation from  $H$-space  to the eigenvalue and eigenvector space. The presence of complicated matrix constraints for the cases $2,3,4$ however makes it  determination technically difficult. This motivates us to consider an alternative route, i.e. numerical investigation of the statistical properties of the ensembles mentioned in previous section. 


To take account of the matrix constraints in each cases stated above, the ensembles are numerically generated as follows. For case 1,  $C$ is obtained by choosing matrix elements such that its first row  consists of the elements  $C_{1n}$, with $n=2\rightarrow N$ independent of each other; the remaining  matrix elements are given by the circulant as well as column constraints with $\alpha=0$. In the case 2, the elements of the first row and and first column, i.e., $C_{1n}$ and $C_{n1}$ respectively with $n=1\rightarrow N$, are chosen to be random and other elements are obtained by invoking  Toeplitz constraint. For the case 3, all $C_{1n}$ ($n=2\rightarrow N$) are chosen from a Gaussian distribution with $C_{11}$ subjected to row constant $\alpha=0$. This is followed by setting all the diagonals $C_{kk}=C_{11}$ for $k=2,3,...,N$. All the off-diagonals of $C$ are chosen randomly except one element from each rows and columns which are determined by the column-sum rule. In the case 4, the first row of $C$ is chosen to have elements $C_{1n}$ (with $n=1\rightarrow N$) as random and the elements in the upper triangle are restricted as $C_{k,k+l}=C_{1,l+1}$ for $l=1\rightarrow (N-1)$ with the elements in the lower triangle to be independent of each other. For case 5, all the elements of block matrix $C$ are chosen to be random except the last column $C_{mN}$ and the last row $C_{Nm}$ (where $m=1\rightarrow N$); the latter ones are obtained by invoking column sum rule (with $\alpha=0$). 

Our next step is to numerically compute the eigenvalues and eigenfunctions of $H$ for each of the five ensembles. This is achieved by using standard LAPACK subroutines for exact diagonalization of the Hermitian matrices. An analysis of the spectral fluctuations requires a prior knowledge of average of the spectral density which can be obtained by averaging over an ensemble, over an spectral range or both. The correct averaging procedure  however depends on the ergodic nature of the spectrum \cite{bg}. This can be explained as follows. For complex systems, the density of states $\rho_e$ at an energy $e$, defined as $\rho_e(e)= \sum_{k=1}^N \delta(e-e_k)$, can often be expressed as a superposition of the fluctuations over an average smooth background: $\rho_e(e) = \rho_{sm}(e) + \delta \rho_e$. Here $\rho_{sm}$ refers to the spectral average at $e$, defined as $\rho_{sm} = \frac{1}{\Delta e} \int_{e-\Delta e/2}^{e+\Delta e/2} {\rm d} e \; \rho_e(e)$, over a scale larger than that of fluctuations, i.e., $ \int_{e-\Delta e/2}^{e+\Delta e/2} {\rm d} e \; \rho_{\rm fluc}(e) =0$. For comparison of the fluctuations at an energy $e$, therefore, it is  necessary to first rescale each spectrum so as to have a same mean level density. This requires a prior information about $\rho_{sm}(e)$. In case of an ergodic spectrum, $\rho_{sm}$ can however be replaced by $R_1(e)$, the ensemble averaged level density (the ergodicity condition of $\rho_e$ is defined as $\langle \rho_{sm}(e) \rangle= R_1(e)$, where $\langle.\rangle$ is an ensemble average for a fixed $e$ \cite{bg}). 

Figure 1 compares the $R_1(e)$ as well as $\langle \rho_{sm}(e) \rangle$ for an ensemble of matrices of fixed size ($2N=1000$) for each of the five cases. Here in case 3, a large deviation between the two curves confirms non-ergodic behavior whereas for other cases deviation is small although not negligible away from $e=0$ region. The figure also reveals a drastic change of level-density from one case to the other, clearly indicating its strong sensitivity to the number of constraints. Interestingly however, as displayed in figure 2,  the rescaled $\langle \rho_{sm}(e) \rangle$ remains size-independent for all the five cases (with rescaling $e\rightarrow e/\sqrt{2N}, \langle \rho_{sm}\rangle \rightarrow \langle \rho_{sm}\rangle \times 2N$), thus implying a same $N$-dependence for all of them.

Due to lack of ergodicity, we unfold the spectrum in each case by the local unfolding process \cite{un}; (the later is based on first  obtaining a smoothed histogram of $\rho_{sm}$ for each spectra (i.e for each matrix) followed by a numerical averaging (i.e. $r_n=\int_{-\infty}^{e_N} \rho_{sm} \; {\rm d}e$). Here we consider the local fluctuations for both high and low density regions of the spectrum and  choose an optimized range $\Delta e$ ($5\%$ of the total eigenvalues), sufficiently large for the good statistics with minimum mixing of different statistics from  the energy range $e \sim (-0.75\pm 0.05)\times \sqrt{2N}$  (bulk) and $e \sim (0\pm 0.03)\times \sqrt{2N}$ (center). This gives approximately $2.5\times 10^5$ eigenvalues for each ensemble of $ 850 $ matrices of size $N=5832$. It is worth noting here that although the density in the bulk region is locally stationary, there is a rapid variation of $\rho_{sm}$ in the center for the cases with reduced number of independent parameters. 
Hence for comparison in the center, it is necessary to choose levels within smaller spectral ranges. The statistics can however be improved by applying ensemble average along with spectral average.

For fluctuations-analysis, we consider two spectral  measures namely
the nearest-neighbor spacing distribution $P(s)$ and the number-variance $\Sigma^2(r)$, the standard tools for the short and long-range spectral correlations, respectively \cite{fh, brody}. Here $P(s)$ is defined as the probability of two nearest neighbour eigenvalues to occur at a distance $s$,  measured in units of local mean level spacing $D$, and $\Sigma^2(r)$ as the variance in the number of levels in an interval of length $r$ mean level spacings. As indicated by previous 
studies, the level fluctuations of a system subjected only to Hermitian constraint along with time-reversal symmetry in a fully delocalized wave limit behave similar 
to that of a Gaussian orthogonal ensemble (GOE)\cite{me,fh,psijmp} with $P(s)= {\pi\over 2} \; s \; {\rm e}^{-{\pi\over 4} s^2}$ and $\Sigma^2(r) \approx {2\over \pi^2} \; \left(\ln (2 \pi r) + \gamma+1-{\pi^2 \over 8} \right)$ with $\gamma=0.5772$. Similarly the fully localized case shows a behavior typical of a set of uncorrelated random levels, that is, exponential decay for $P(s)$, also referred as Poisson distribution, $P(s) = {\rm e}^{-s}$, and $\Sigma^2(r) = r$ \cite{me,fh,psijmp}. But, as discussed in \cite{tss} for the case of chiral circulant matrices, Poisson spectral statistics appears along with delocalized eigenfunctions. This indicates the influence of 
constraints on the relation between eigenvalue and eigenfunction statistics which is further confirmed by the present study of other four cases.    

Figure 3 compares $P(s)$ and $\Sigma^2(r)$ of the spectra for all five cases at two different energy ranges. As displayed in the figure, the level statistics changes from Poisson to GOE as the number of independent matrix parameters increase. As expected on theoretical grounds \cite{ss2,tss}, Case 1 and 5 show  Poisson and almost GOE like behavior respectively for all energy ranges. For cases 2 and 4, $P(s)$ is intermediate between Poisson and GOE (though they differ from each other depending on the number of independent matrix elements) in bulk $(e \sim -0.75\pm 0.05)$ as well as near center $(e \sim 0\pm 0.03)$ implying a partial localization of eigenfunctions. For the case 3, although $P(s)$ in the bulk is close to GOE  (Fig 3a) however the deviation of its statistics from GOE is reflected in large $r$ behavior of $\Sigma^2(r)$ which   is intermediate between GOE and Poisson (Fig 3b).  Similarly, at the center of case 3, $P(s)$ is intermediate to Poisson and GOE (Fig 3c) although $\Sigma^2(r)$ approaches Poisson limit (Fig 3d). (Note, due to rapidly changing density  in this case,  $\Sigma^2(r)$ statistics for large $r$ is more susceptible to  unfolding-issues near $e=0$. A short range statistics e.g. $P(s)$ is therefore a more reliable criteria of the fluctuations in this case).

The variation of statistics with energy for case 3, from almost GOE type behavior to an intermediate state between Poisson and GOE, suggests an existence of  a mobility edge separating extended states from partially localized ones. To seek criticality for case 3 at bulk as well as at the center of the spectrum, we analyzed the behavior of $P(s)$ and $\Sigma^2(r)$ for many matrix sizes. Figure 4  confirms the size-independence of $P(s)$ around $e=0$ (fig 4c), thus implying a critical spectral statistics different from both GOE and Poisson even in infinite-size limit. Although the behavior of  number variance $\Sigma_2(r)$ does indicate a size-dependence for large $r$, this however does not rule out criticality. This is because  the number variance, requiring averaging over large spectral ranges is not an appropriate measure in this case near $e\sim 0$. The criticality for this case is however confirmed later by an eigenfunction statistical measure too.

As mentioned in section I, one of our primary objectives is to understand the influence of matrix  constraints on the eigenfunction dynamics. For this purpose, we consider  the inverse participation ratio (IPR) $I_2$, a standard tool to describe the localization behavior. For an eigenfunction $O_n$, with $O_{kn}$ as its components corresponding to an eigenvalue $e_n$, 
it is defined as $I_2(O_n)= \sum_{k=1}^N  |O_{kn}|^4$. In general, IPR  varies with energy and it is a standard practice to consider the average of IPR ($I_{2}$) of all eigenfunctions within a given spectral range in which the average spectral density varies smoothly.  But as discussed above, the latter shows a rapid variation for some specific energy ranges i.e $e=0$ as well as near edges and  it is more appropriate to  consider the ensemble averaged $I_{2}$, referred as $\langle I_{2}\rangle$ at a specific energy instead of the spectral averaged one.

As displayed in Fig 5, almost all  eigenfunctions are delocalized with $\langle I_2 \rangle \approx {3 \over 2N} $ for all five cases. Further,  for the cases with column-constraint (i.e cases 1, 3 and 5), $\langle I_2 \rangle$ for the eigenfunction corresponding to eigenvalue $\alpha =0$ is $\approx 1/N$, with $\alpha$ as the column constant (see Fig 5(a), 5(c) and 5(e)). This indicates  an extended eigenfunction  statistics  in the basis-space for all the five cases, irrespective of the number of constraints. Clearly the eigenfunction statistics for these cases is sensitive only to strength of the disorder. As the latter is chosen same for all the independent elements, this results in extended dynamics in the basis space. 

Another point worth indicating here is the following. For the cases 3 and 5, the largest eigenvaule pairs are isolated lying quite far away from the bulk and the $\langle I_2 \rangle$ for corresponding eigenfunctions are much larger indicating their localization. Fig 6 shows that for case 3 (Fig 2a), there is  a  ''pairwise localization" corresponding to largest eigenvalue pair $e_1, e_{N+1}$: $I_2(O_1)=I_2(O_{N+1})=1/2$. The latter is a characteristic of column-row constraint matrix \cite{ss1} whereas in case 5 there are two pairs of such localized eigenfunctions (Fig 6b).

As clear from Fig 5, typically $\langle I_2 \rangle \propto {1 \over 2N} $ for all energy ranges, indicating that a  typical eigenfunction in each case tends to delocalize itself in the basis space. This is in contrast with the spectral statistics which undergoes  variation from Poisson to GOE with decreasing number of matrix constraints. It must be noted that  the Poisson and GOE statistics of the eigenvalues are usually believed, respectively, to be an indicator of the localized and delocalized dynamics of eigenfunctions in the basis space. But as our analysis clearly indicates this is not the case for the chiral ensembles of structured matrices.

To confirm the critical behavior of case 3, we analyze  the correlation(fractal) dimension ($D_q$), a frequently used characteristic of the eigenfunction localization, defined as $D_q = -\frac{\langle lnI_q\rangle}{ln N}$. For localized eigenfunction, $D_q =0$ whereas its value increases to system dimension$d$ as localization decreases. If $0>D_q>d$, this is an indicator of the multifractality of the eigenfunction statistics. As shown in Fig 7(a), (b) and (c), $D_q$ is almost size-independent  everywhere in the spectrum. Fig 7(d) however indicates an energy-dependence of $D_q$: it varies from its near localized limit ($D_q \sim 0.1$) near spectrum edge to a partial localized value near some intermediate energy ($D_q \sim 0.6$)  to extended limit ($D_q \sim 0.9$) near center. This suggest a weak-multifractality of the eigenfunctions near the center.

\section{Conclusion}

Based on the present study of a few structured matrix ensembles, we believe that the relation between the statistical behavior of the eigenvalues and eigenfunctions  of a generic random matrix ensemble is lot more complicated than previously believed on the basis of Hermitian ensembles with symmetry as the only constraint. In fact it seems while the spectral statistics is primarily governed by the number of independent matrix parameters, the eigenvector statistics is sensitive to their relative degree of randomness. More clearly, as far as the independent matrix elements are statistically of the same strength (e.g same mean and variance) irrespective of their number, it will always lead to   almost all extended eigenfunctions (except for a very few strongly localized ones). The eigenvalue statistics however undergoes significant change as the number of the independent parameters are varied. 

It must be emphasized here that while the results obtained in this work are based on the numerical analysis, the behavior for case 1 and case 5  can also be explained on theoretical grounds (see \cite{tss} and \cite{ss1} respectively). A theoretical understanding of cases 2, 3, 4 and in general of a generic ensemble subjected to a set of matrix constraints is although very desirable but is technically complicated.  We expect to report  our ongoing attempts in this context in near future.



\newpage

\begin{figure}
\centering
\includegraphics[width=1.2\textwidth,height=1.0\textwidth]{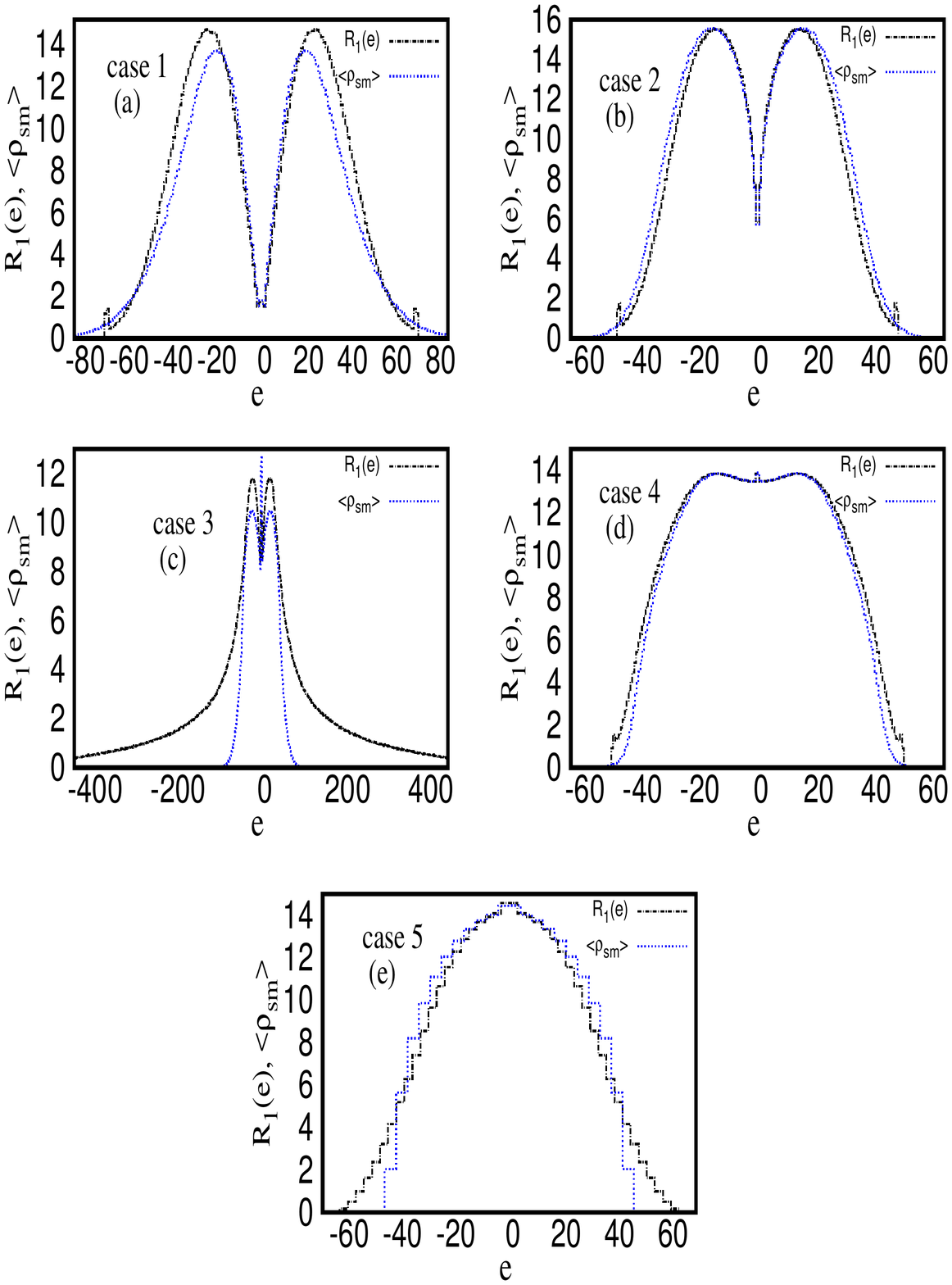} 
\vspace*{-15 mm}
\caption{ 
\textbf{Ergodicity in level density:}
Comparison between ensemble averaged level density $R_1(e)$ and $\langle \rho_{sm}(e) \rangle$ for matrix $H$ of fixed size $2N=1000$ depicts nearly ergodic level density ($R_1(e)=\langle \rho_{sm}(e) \rangle$) around center ($e\sim 0$) energy regime for all the cases except case 3. Fig(c) clearly potrays non-ergodicity at any energy range for case 3.}
\label{fig1}
\end{figure}

\begin{figure}
\centering
\includegraphics[width=1.2\textwidth,height=1.0\textwidth]{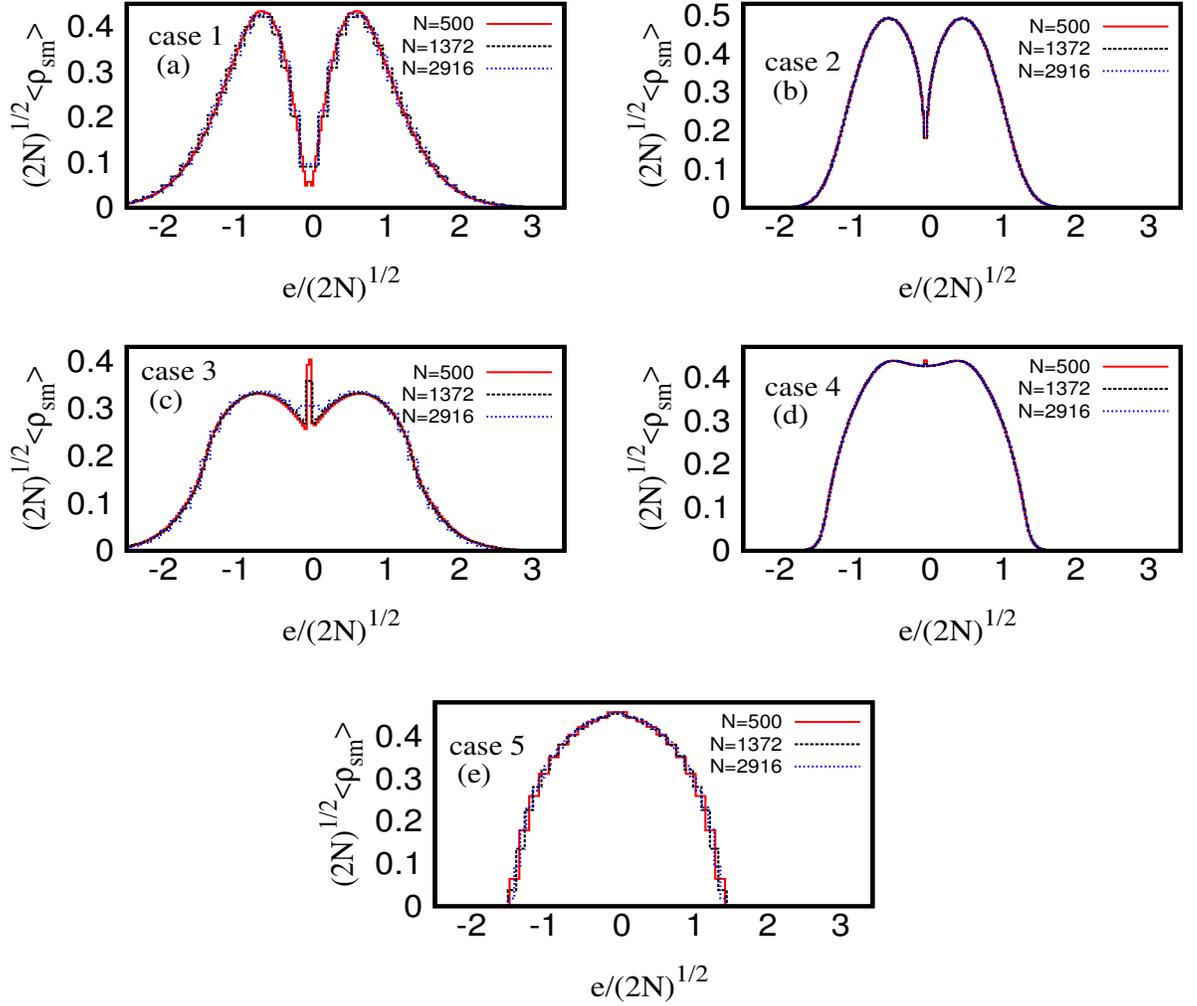} 
\vspace*{-15 mm}
\caption{ 
\textbf{Size-dependence of level density:}
This figure depicts the size-dependence of the density of states for the five cases. After the rescaling of the axes: $e\rightarrow e/\sqrt{2N} , \rho_{sm}\rightarrow \rho_{sm}\times \sqrt{2N}$, the $\langle \rho_{sm}(e) \rangle$ for different matrix size ($2N$) superpose with each others.}
\label{fig2}
\end{figure}

\begin{figure}
\centering
\includegraphics[width=1.2\textwidth, height=1.2\textwidth]{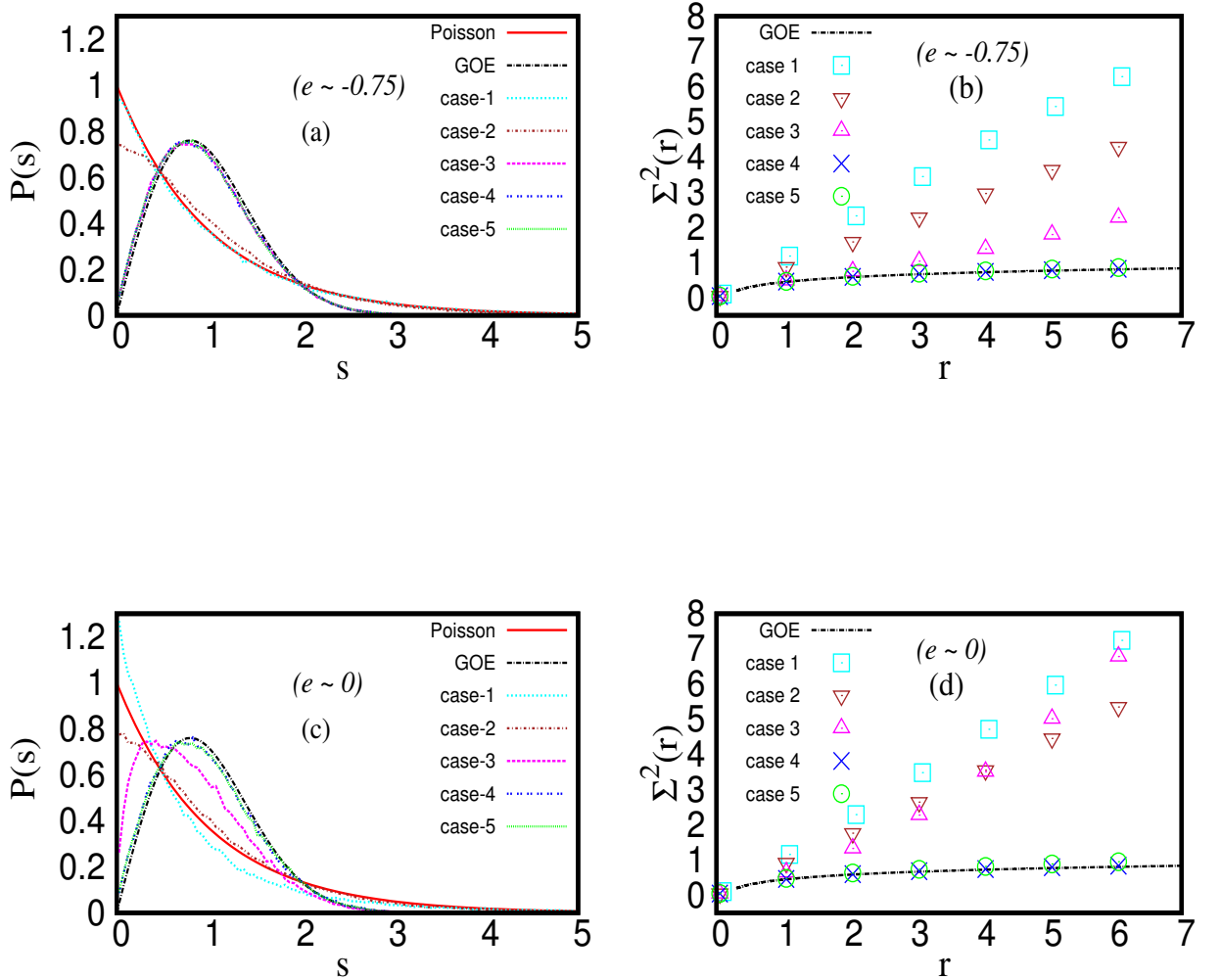} 
\vspace*{-30 mm}
\caption{
{\bf Spectral statistics at two different energy regime}: Fig (a) and (b) show $P(s)$ and $\Sigma^2(r)$ at bulk ($e \sim (-0.75\pm 0.05)\times \sqrt{2N}$) and (c) and (d) are at center ($e \sim (0\pm 0.03)\times \sqrt{2N}$) energy regime. Case 1 and 5 cleary shows Poisson and almost GOE like behavior respectively whereas case 2 approaches to Poisson and case 4 tends to GOE statistics. But for case 3, although $P(s)$ at bulk(fig (a)) approaches to GOE but at center $P(s)$ is intermediate between GOE and Poisson statistics(fig(c)) and $\Sigma^2(r)$ at bulk shows intermediate statistics(fig(b)) but at center it approaches to Poisson statistics(fig(d)). This kind of behavior indicates criticality in case 3. These plots are for matrix of size $2N=5832$.}
\label{fig3}
\end{figure}

\begin{figure}
\centering
\includegraphics[width=1.2\textwidth, height=1.2\textwidth]{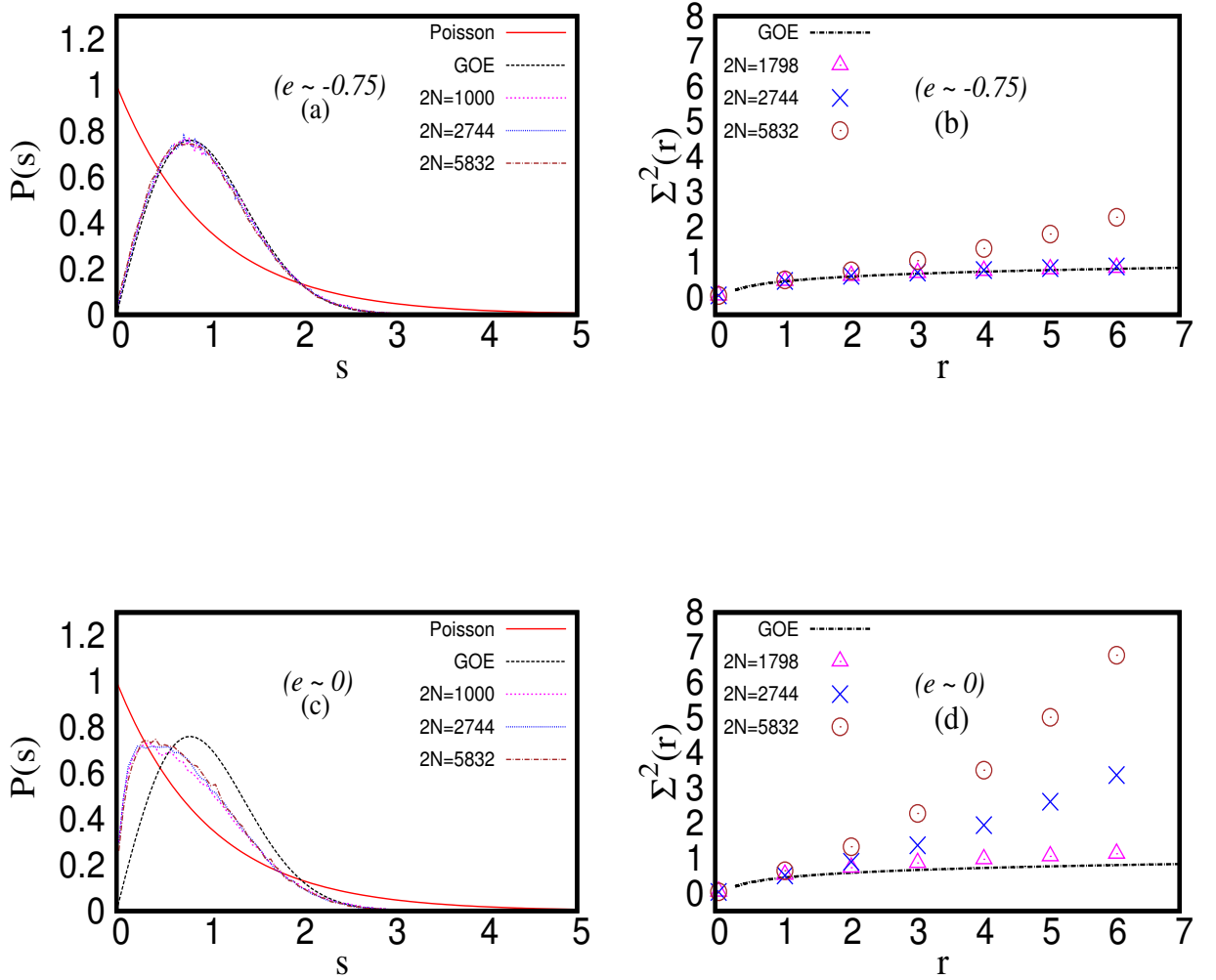} 
\vspace*{-30 mm}
\caption{
{\bf Criticality for case 3 at bulk and center of the spectrum}: At bulk ($e \sim (-0.75\pm 0.05)\times \sqrt{2N}$), we did not find criticality in spectral statistics  (fig (a) and (b)). But Fig (c) shows a critical behavior of $P(s)$ at center ($e \sim (0\pm 0.03)\times \sqrt{2N}$) as it is independent of the size of the matrix but $\Sigma^2(r)$ does not show criticality anywhere in the spectrum (fig (b) and (d)).}
\label{fig4}
\end{figure}

\begin{figure}
\centering
\includegraphics[width=1.2\textwidth, height=1.2\textwidth]{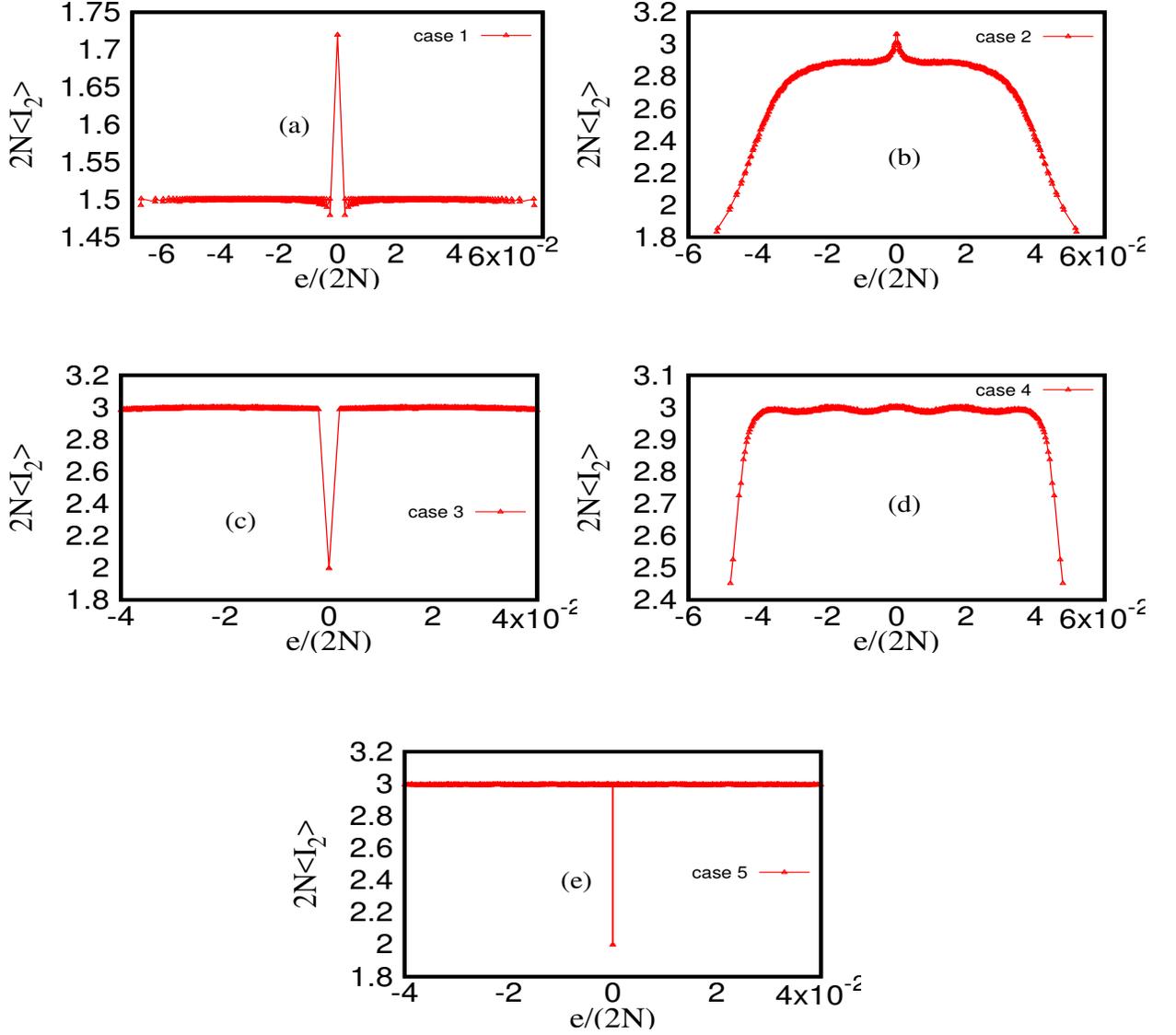} 
\vspace*{-22mm}
\caption{
\textbf{Extended eigenfunction statistics}: This figure shows that ensemble averaged inverse participation ratio $\langle I_2 \rangle$ for all the five cases are $\sim \frac{3}{2N}$ implying delocalized eigenfunctions statistics for all expect the largest pairs of eigenvalues for case 3 and 5. Note that fig (c) and (e) are plotted without the largest pairs of eigenvalues (see Fig 2). For the cases where column-constraint($\alpha$) is there (i.e., case 1, 3 and 5), $\langle I_2 \rangle$ corresponds to the eigenvalue $e_n=\alpha=0$ is $1/N$. The size-analogy follows on the rescaling $e \rightarrow e /(2 N)$ , $\langle I_2\rangle \rightarrow \langle I_2 \rangle\times (2 N)$. All of them are for an ensemble of $5000$ matrices of a fixed size $2N=1000$.}
\label{fig5}
\end{figure}

\begin{figure}
\centering
\includegraphics[width=1.2\textwidth, height=1.2\textwidth]{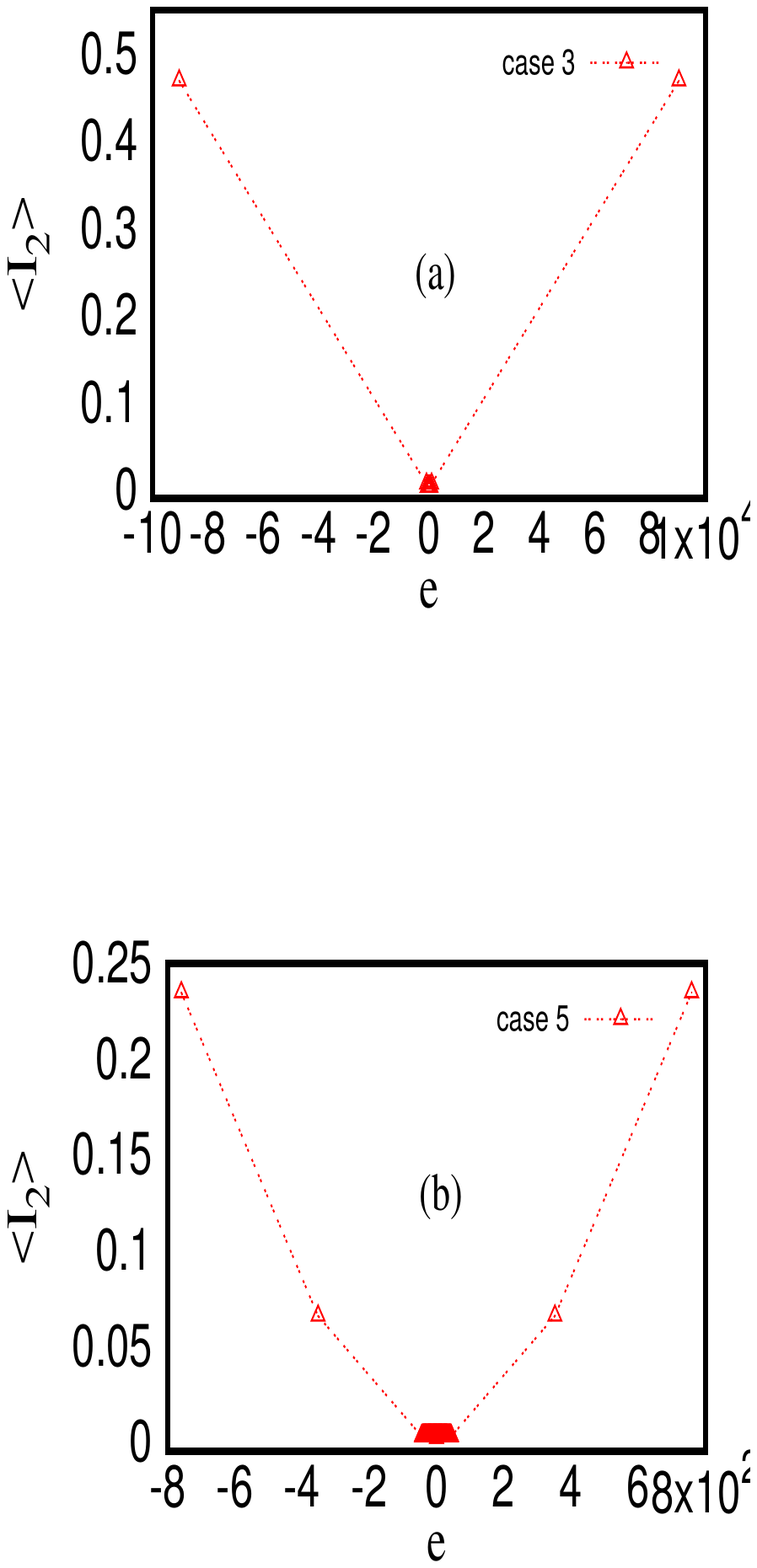} 
\vspace*{-22mm}
\caption{
\textbf{Localization in eigenfunctions}: This figure shows that, for case 3 and 5, $\langle I_2 \rangle$ corresponding to largest pairs of eigenfunctions are much higher than the bulk indicating localizations in eigenfunction statistics. for an ensemble of $5000$ matrices of a fixed size $2N=1000$ with Gaussian disorder. For case 3 (fig 2a), corresponding to largest eigenvalue pair, $\langle I_2 \rangle = 1/2$  but for case 5 (fig b), there are two pairs of extreme eigenvalues and corresponding $\langle I_2 \rangle = 1/4$, and $\approx 1/20$.  }
\label{fig6}
\end{figure}

\begin{figure}
\centering
\includegraphics[width=1.2\textwidth, height=1.2\textwidth]{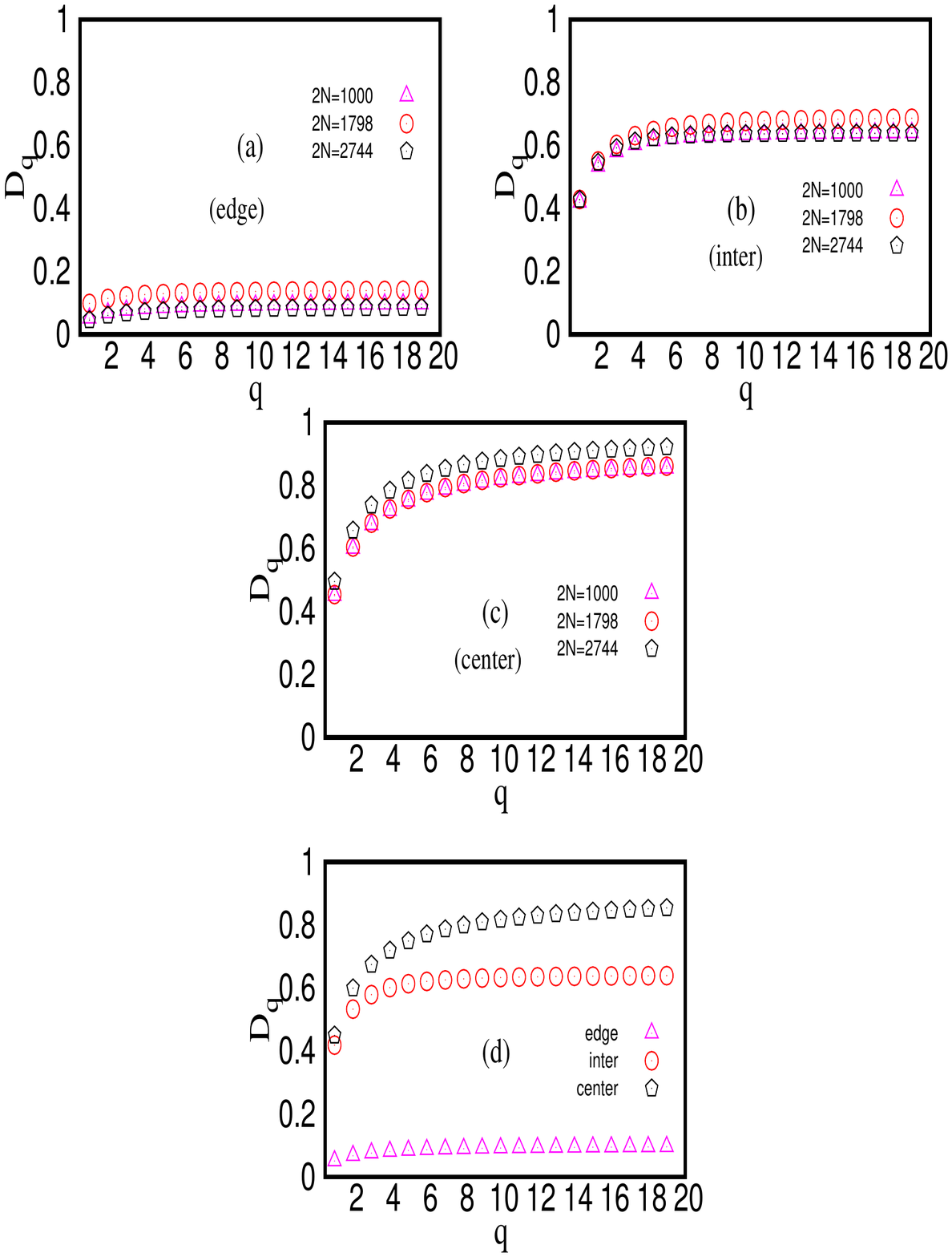} 
\vspace*{-22mm}
\caption{
\textbf{Fractal dimension $\boldsymbol {D_q}$ for case 3}: Figure (a), (b) and (c) indicate critical behavior corresponding to different energy levels, i.e., \textit{edge} ($e$ = extreme negative eigenvalue), \textit{inter} ($e$ is somewhere in negative x-axis greater than extreme eigenvalue) and \textit{center} ($e=0$) for case 3 . Fig (d) confirms localization in eigenfunction statistics at \textit{edge}, i.e, eigenfunction correspond to largest eigenvalue is localized for case 3, otherwise eigenfunction statistics is extended. Fig (d) is plotted for matrix of size $2N=1000$.}
\label{fig7}
\end{figure}


\newpage

\begin{thebibliography}{10}

\bibitem{psijmp}
P. Shukla, Int. J. Mod. Phys. B (WSPC) 26, 12300008, (2012). 

\bibitem{fh}
F. Haake,{\it  Quantum Signatures of Chaos}, Springer (Berlin), (1991). 

\bibitem{me}
M. L. Mehta, {\it  Random Matrices}, Academic Press, (1991). 

\bibitem{brody} T.A.Brody, J.Flores, J.B.French, P.A.Mello, A.Pandey and 
S.S.M. Wong, Rev. Mod. Phys. 53, 385, (1981). 

\bibitem{dy}
F.Dyson, J. Math. Phys. 3, 1191 (1962).

\bibitem{zirn}
A. Altland and M.R.Zirnbauer, Phys. Rev. B 55, 1142, (1997). 

\bibitem{qw} R. Qiu and M.Wicks, Cognitive Networked Sensing and Big Data, 1st ed. (Springer, New York, 2013).

\bibitem{ach} D. Achlioptas, Random matrices in data analysis, in Knowledge Discovery in 
Databases: PKDD 2004 (springer-Verlag, Berlin, Heridelberg, 2004). P.1.




\bibitem{cd} R. Couillet and M. Debbah, Random Matrix Method for Wireless Communications, 1st ed. (Cambridge University Press, Cambridge, 2014).

\bibitem{so} C. Soize, Journal of Sound and Vibration 263, 893 (2003).

\bibitem{grela} J. Grela and T. Guhr, Phys. Rev. E 94, 042130 (2016).

\bibitem{afm} Y. Ahmadian, F. Fumarola and K.D. Miller, Phys. Rev. E 91, 012820 (2015).

\bibitem{ahn}
A. Amir, N. Hatano and D. Nelson, Phys. Rev. E 93, 042310 (2016).

\bibitem{by} P. Bourgade and H.T.Yau, Commun. Math. Phys. 350, 231, (2017).

\bibitem{ossi} K. Truong and A. Ossipov, arXiv: 1708.05345v1.

\bibitem{chal}
V. Gurarie and J.T. Chalker, Phys. Rev. B, 68, 134207, (2003).


\bibitem{mm} M. Mezard, G. Parisi and A. Zee, Nucl. Phys. B, 559, 689, (1999);
J. Staring, B. Mehlig, Y.V.Fyodorov and J. M. Luck, Phys. Rev. E, 67, 047101, (2003);
T.S. Grigera , V. Martin-Mayor, G. Parisi and P. Verrocchio, Phys. rev. Lett. 87, 085502, (2001); J. Phys. Cond. Mat. 14, 2167 (2002); Nature 422, 289 (2003); M. Rusek, J. Mostowski and A. Orlowski, Phys. Rev. A, 61, 022704 (2000); F. A. Pinheiro, M. Rusek, A. Orlowski, and B. A. van Tiggelen, Phys. Rev. E 69, 026605, (2004); M. Antezza, Y. Castin and D. A. W Hutchinson, Phys. Rev. A, 82, 043602, (2010); S. Skipetrov and R. Maynard, Phys. Rev. Lett. 85, 736 (2000);  B. Zyuzin and A. Spivak, Phys. Rev. Lett. 84, 1970 (2000);  B. Gremaud and T. Wellens, Phys. Rev. Lett., 104, 133901 (2010).

\bibitem{ggs} B. Georgeot, O. Giraud, D.L. Shepelyansky, Phys. Rev. E 81, 056109, (2010).

\bibitem{cmz}
D. Challet and M. Marsilli, R. Zecchina, Phys. Rev. E, 84, 1824, (2000).

\bibitem{zlwlty}
L. Zhao, S. Liao, Y. Wang, Z. Li, J. Tang, B. Yuan, arXiv: 1703.001v44 [cs.LG].

\bibitem{pstrans}
P. Shukla and I. Batra, Phys. Rev. B, (2005).

\bibitem{psand}
P.Shukla, Phys. Rev. E, 62, 2098, (2000);
J.Phys.: Condens. Matter 17, 1653, (2005).

\bibitem{evan}
S N Evangelou and D E Katsanos, J. Phys. A: Math. Gen.36 (2003) 3237–3254 

\bibitem{tss} T. Mondal, S. Sadhukhan, and P. Shukla, Phys. Rev. E: 95, 062102 (2017) 

\bibitem{ss1} P. Shukla, S. Sadhukhan, J. Phys. A: 48 (2015) 415002.

\bibitem{ss2} S. Sadhukhan, P. Shukla, J. Phys. A: 48 (2015) 415003.



\bibitem{toep} R. M. Gray,  Foundations and Trends® in Communications and Information Theory: Vol. 2, No. 3, pp 155-239 (2006)

\bibitem{bg} O. Bohigas and M.J.Giannoni, Ann. Phys. 89, 422, (1975).

\bibitem{un}
J. M. G. Gomez, R. A. Molina, A. Relaño, and J. Retamosa
Phys. Rev. E 66, 036209

 


 



\end{thebibliography}
\end{document}